\documentclass[10pt,twocolumn]{article}
\usepackage{geometry}
\usepackage[absolute,showboxes]{textpos}

\TPMargin{5pt}

\newcommand{\copyrightstatement}{
    \begin{textblock}{0.84}(0.08,0.93)    
         \noindent
         \footnotesize
         \copyright2020 IEEE. Personal use of this material is permitted. Permission from IEEE must be
obtained for all other uses, in any current or future media, including
reprinting/republishing this material for advertising or promotional purposes, creating new
collective works, for resale or redistribution to servers or lists, or reuse of any copyrighted
component of this work in other works.
    \end{textblock}
}

\usepackage{cite}
\usepackage{amsmath,amssymb,amsfonts}
\usepackage{algorithmic}
\usepackage{graphicx}
\usepackage{textcomp}
\usepackage{authblk}
\usepackage{float}
\usepackage[shortlabels]{enumitem}

\begin{document}

\copyrightstatement

\title{The distribution of inhibitory neurons in the \textit{C. elegans} connectome facilitates self-optimization of coordinated neural activity
}

\author[1]{Alejandro Morales}
\author[2]{Tom Froese}
\affil[1]{National Autonomous University of Mexico\\
alejandroe@ciencias.unam.mx}
\affil[2]{Okinawa Institute of Science and Technology Graduate University\\
tom.froese@oist.jp
}

\date{}

\maketitle

\begin{abstract}
The nervous system of the nematode soil worm \textit{Caenorhabditis elegans} exhibits remarkable complexity despite the worm's small size. A general challenge is to better understand the relationship between neural organization and neural activity at the system level, including the functional roles of inhibitory connections. Here we implemented an abstract simulation model of the \textit{C. elegans} connectome that approximates the neurotransmitter identity of each neuron, and we explored the functional role of these physiological differences for neural activity. In particular, we created a Hopfield neural network in which all of the worm's neurons characterized by inhibitory neurotransmitters are assigned inhibitory outgoing connections. Then, we created a control condition in which the same number of inhibitory connections are arbitrarily distributed across the network. A comparison of these two conditions revealed that the biological distribution of inhibitory connections facilitates the self-optimization of coordinated neural activity compared with an arbitrary distribution of inhibitory connections. 

\end{abstract}

\thispagestyle{empty}

\section{Introduction}

The brain has been called the most complex object in the universe. It is therefore understandable that we are still lacking a unified theory of how the brain works. The various mechanisms underlying its capacity to generate ordered patterns of neural activity in a spontaneous manner are still being unravelled. Unlike artificially designed systems, like the digital computer, the brain is organized in a heterarchical manner and neural activity is not governed by a central controller. This inherent complexity poses a challenge for rapid progress for brain science, but at least theoretical neuroscience can benefit from general insights coming from recent advances in the theory of complex adaptive systems.

A relatively unknown but promising approach is the study of self-optimization using an algorithm first proposed by Watson et al. \cite{watson2011global} on the basis of the Hopfield neural network, and which was subsequently developed in various directions. The core idea is that a Hopfield network can be made to form an associative memory of its own state attractors, namely by repeatedly letting it minimize neural constraints posed by the network architecture and reinforcing those attractors on which it converges. The result is that the network starts to generalize over its local attractors and thereby enhances its capacity to converge on more optimal attractors. In contrast to much other contemporary work in training artificial neural networks, this self-optimization process does not rely on any prior knowledge of the target domain; it is based on a simple form of unsupervised learning (e.g. Hebbian learning). Previous work has successfully applied this algorithm to various kinds of neural network architectures; for example, in networks with symmetric weights \cite{watson2011optimization,watson2011transformations}, networks with continuous activation functions \cite{zarco2018self,zarco2018continous}, and spiking neural networks \cite{woodward2015neural}. 

However, in all previous models of self-optimization the neural networks were relatively small (e.g. around 100 neurons) and had a highly artificial topology (e.g. arbitrary initialization of connection weights, arbitrary distribution of negative connections, or perfectly modular arrangement of connections). This leaves it unclear whether the process would successfully scale up to the complexity of biological neural network architectures. In other words, if one of the aims of this particular model of a complex adaptive system is to provide inspiration for theoretical neuroscience, then it would be more meaningful to test this algorithm with a biologically plausible neural network architecture. 

Accordingly, as a first step in this direction, we proposed that the \textit{C. elegans} connectome is a suitable starting point \cite{morales2019alife}. \textit{Caenorhabditis elegans} is a one-millimeter-long soil worm of the nematode family. It is the first animal from which the complete mapping of synaptic connections has been obtained and it is a reference animal model in biology \cite{white1986structure,worm2006review}. The neural network dynamics of \textit{C. elegans} have been modeled in several ways \cite{izquierdo2018role}, and the artificial life research community has also shown an interest in the study of \textit{C. elegans} \cite{winkler2009,hattori2012modeling,beer2016propagation,aguilera2017,dahlberg2020}. However, it has focused on small circuits that are embodied in a sensorimotor loop, without taking into account the whole connectome.

We took an alternative approach and decided to study the whole connectome as a complex adaptive system. We recently demonstrated that the self-optimization algorithm can be applied to the \textit{C. elegans} connectome when it is viewed as a directed multigraph \cite{morales2019alife}. Some previous work on self-optimization employed networks without negative connections, so as a first step we explored the effect of including negative connections in the connectome. In that work we compared two conditions: (1) The self-optimization algorithm was run with only excitatory (positive) connections, and (2) it was run with 30\% inhibitory (negative) connections that were arbitrarily assigned at the beginning of the algorithm. We found that the \textit{C. elegans} connectome consistently showed a tendency to optimize its own connectivity such that neural coordination become more prominent, but the presence of negative connections tended to increase the difficulty of neural coordination.

Given that this difficulty may result from the arbitrariness of the distribution of the negative connections, we next explored self-optimization in the \textit{C. elegans} connectome by separating it into hierarchically organized functional clusters that are known from the literature \cite{sohn2011,morales2020unsupervised}. In that work, self-optimization was tested in two scenarios: (1) In each cluster independently from the rest of the connectome, and (2) in each cluster but embedded within the context of the whole connectome. Inhibitory connections were introduced arbitrarily in (1) and distributed in inter-cluster connections in (2). Again, the \textit{C. elegans} connectome consistently showed a tendency to optimize its own connectivity in both cases, but there was a significant improvement of self-optimization in (2). This finding revealed that it is not necessarily the presence of inhibitory connections per se that makes self-optimization less effective, but rather the precise distribution of the inhibitory connections and how they carve up the overall connectome into distinct clusters.

In this paper, we continued exploring the conditions for self-optimization in the \textit{C. elegans} connectome by including more biological details in our model. In particular, we took into account the biological distribution of inhibitory connections as specified by the neurotransmitter identity of each of the neurons in the connectome \cite{pereira2015,riddle97neurotr} (scenario A), and we compared this with a controlbcondition in which the same number of inhibitory connections are distributed arbitrarily across the network (scenario B). 

\begin{figure}[hbt!]
\centering
\includegraphics[width=2.7in]{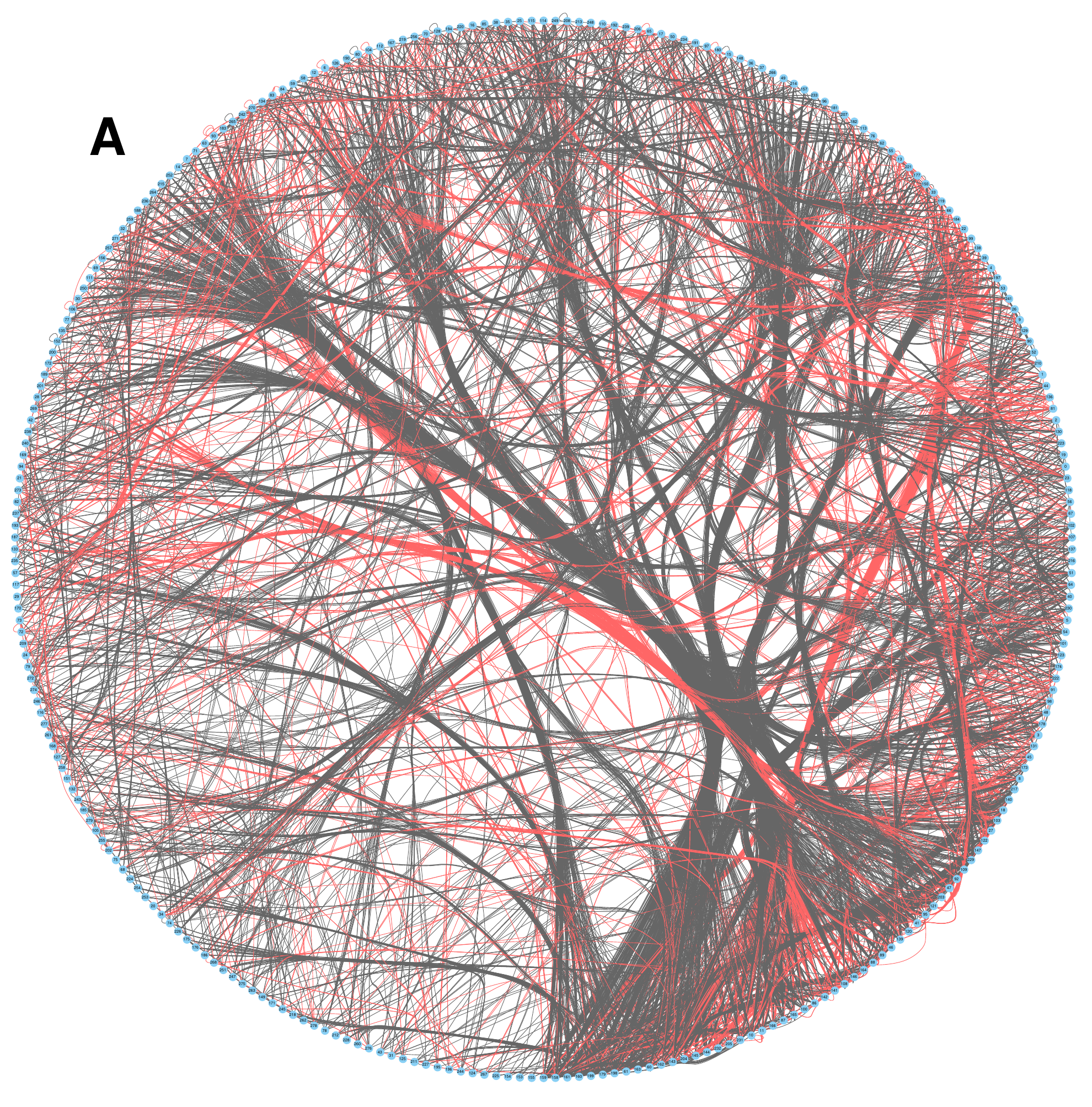}
\vskip 0.6cm
\includegraphics[width=2.7in]{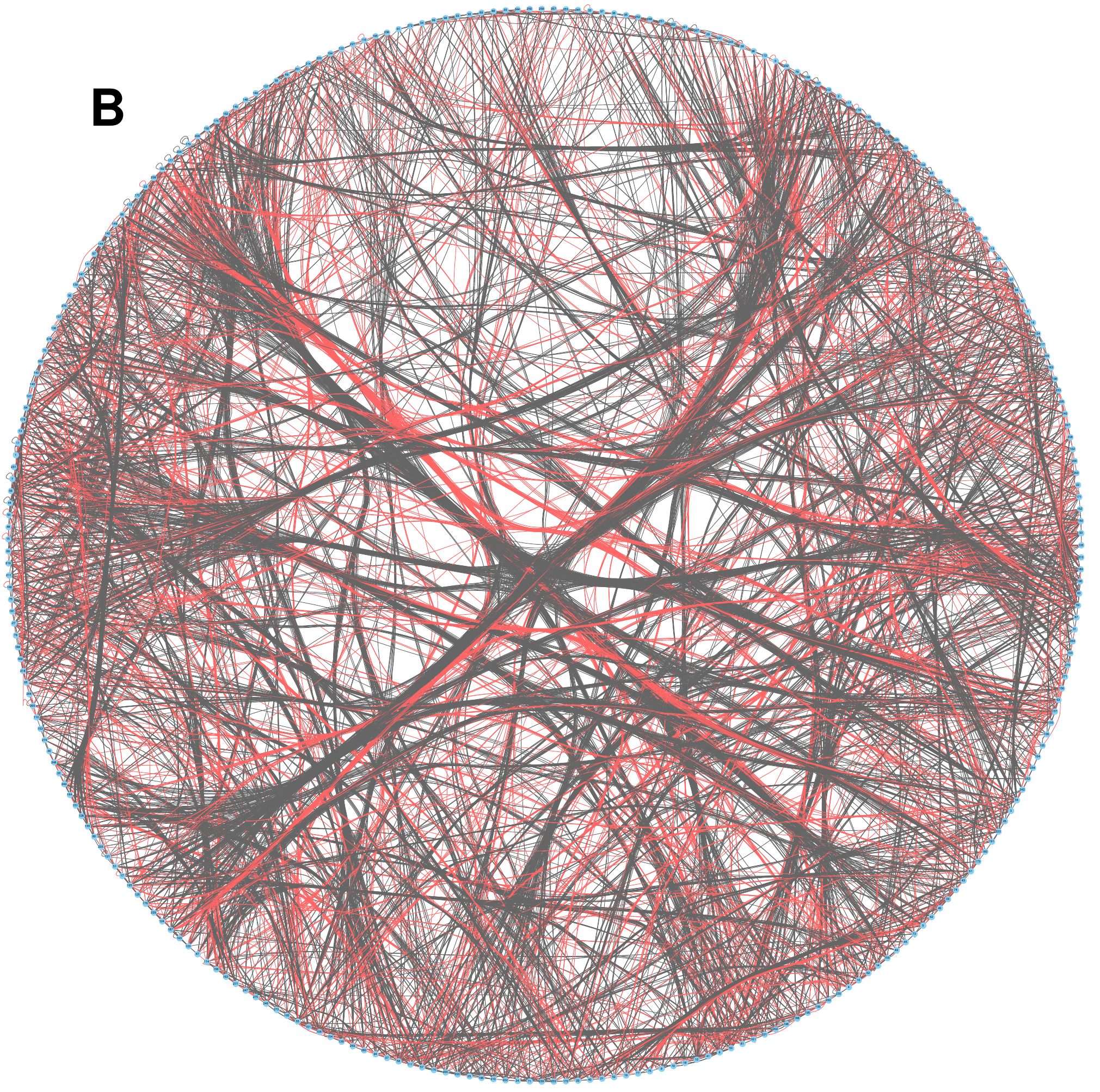}
\caption{\small{Visualization of the \textit{C. elegans} connectome in a circular layout with a bundling procedure. Black represents excitatory connections; red represents inhibitory connections. Scenario \textbf{A}: Connectome with 26\% inhibitory connections distributed in accordance with each neuron's neurotransmitter identity. Scenario \textbf{B}: Connectome with 26\% inhibitory connections but with an arbitrary distribution. By visual inspection we can observe a greater number of aggregated connections in scenario A, especially more clustering of excitatory connections.}}
\label{conn}
\end{figure}

Based on our previous work, we hypothesized that the biological distribution provides a more effective starting point for the self-optimization of connectivity, consistently leading to improved neural coordination. We were able to confirm this hypothesis. 

But we also found self-optimization to be generally less effective, at least compared to previous variations of the model in which we had initialized the connectome as a fully interconnected network by padding its biologically existing connections with initially non-functional (i.e. zero-weighted) connections that are biologically non-existent. We speculate about the reasons for this difference at the end of the paper.

\section{Methods}

\subsection{The connectome}

We followed the same approach as in our previous work \cite{morales2019alife,morales2020unsupervised} regarding the connectome's translation to a directed multigraph, with neurons as nodes and connections as edges. This approach enabled us to run the self-optimization algorithm on a model of the connectome published by Jarrell et al. \cite{jarrell2012connectome}. That connectome database contains hermaphrodite and male neural information such as connection direction, type of connection (synapse or gap junction), and the number of connections between neurons. We only considered hermaphrodite information because \textit{C. elegans} males, which occur at a low frequency in laboratory strains \cite{anderson2010celegans}, differ slightly in the number, identity, and synaptic wiring in their nervous system \cite{oren2016sex}. 

\textit{C. elegans} consists of only 959 cells, of which 302 are neurons belonging to the nervous system. We took into account 280 neurons, with 5609 connections. We did not consider the 20 pharyngeal neurons because they are treated as belonging to an independent neural system \cite{albertson1976pharynx}. We also removed the neurons \textit{CANR} and \textit{CANL} because they do not have obvious connections and are classified as end organs \cite{cook2019whole}.

We modeled chemical synapses as single-directed links between neurons (for example, $A \rightarrow B$ indicates that neuron $A$ is presynaptic to neuron $B$, and $B$ is postsynaptic to $A$). Gap junctions are modeled as double-linked neurons (if two neurons, $C$ and $D$, have a gap junction between them, then there are two links: $C \rightarrow D$ and $D \rightarrow C$). The number of connections between neurons was translated into the weight of each edge, normalized in the interval $(0, 1)$. Both edges representing a gap junction were assigned the same weight. Weight values vary between 1 and 81 before normalization. There were only 15 edges whose weight values exceeded 44, and we decided to treat them as outliers and clip them to 1 during normalization. This was done to prevent the majority of connections to become comparatively too small in weight. The effect of clipping these outliers is to increase the scope of state-space explorations of the resulting network, allowing it to visit a more diverse set of attractors. Future work could develop a more elegant way of translating the connectome information into edge weights.

\subsection {Inhibitory neurons}

We categorized all outgoing connections as either excitatory or inhibitory based on a neuron's neurotransmitter identity \cite{pereira2015,riddle97neurotr}. In other words, when a neuron's identity is determined to be inhibitory, we made the signs of the weights of its outgoing connections negative. According to \cite{pereira2015}, \textit{C. elegans} neurons can be partitioned into one of four groups based on its neurotransmitter identity, with the following counts:

\begin{enumerate}[(a)]
    \item 147 excitatory neurons
    \item 26 inhibitory neurons
    \item 83 simultaneously excitatory and inhibitory neurons 
    \item 24 neurons not characterized (their neurotransmitter identities are unclear)
\end{enumerate} 

Excitatory neurons are mostly dopaminergic neurons, while inhibitory neurons are GABA-aminergic. Neurons in group (c) have inhibitory and excitatory neurotransmitters simultaneously (glutamate, serotonin, octopamine, tyramine) \cite{sawin2000,sanyal2004}. Molecular analysis of \textit{C. elegans} neuron signaling also includes neuropeptide function, receptors and related genes \cite{worm2006review,riddle97neurotr}, but we are not taking into account these molecules in our current model. Regarding neuron function, most excitatory and inhibitory \textit{C. elegans} neurons are motor neurons. Neurons that have excitatory and inhibitory connections simultaneously are usually mechanosensory neurons and interneurons \cite{cook2019whole}.

We assigned positive and negative outgoing connections to the neurons in group (a) and (b), respectively. Then, for groups (c) and (d) we arbitrarily selected 50\% of the neurons as having all of their outgoing connections negative. Under these conditions, we obtained around 26\% inhibitory connections in the whole network: a percentage similar to the one used in our previous studies \cite{morales2019alife,morales2020unsupervised} and close to the theoretical approximation in literature \cite{buxhoeveden2002,hendry1987,brunel2000,triplett2018}.

In Fig. \ref{conn}, we depict the resulting Hopfield network structure in a circular layout, in which black and red edges indicate excitatory and inhibitory connections, respectively. We applied a bundling algorithm to avoid visual clutter, and to give an indication of possible clustering of connections.

\subsection {The connectome as a complete graph}

In previous work we had added zero-weighted connections to the Hopfield neural network of the connectome in order to turn it into a complete directed graph  \cite{morales2019alife,morales2020unsupervised}. In this way, Hebbian learning could include extra functional connections in the network by changing these weights from zero to non-zero values, and we had found that this facilitates the removal of constraints between interactions during the self-optimization process. In the current work, we first tested the restricted network based on the biological connectome of 5,609 edges in a first set of simulation runs. Then we conducted a second set of simulation runs in which we expanded the network into a complete graph by adding extra zero-weighted edges, which resulted in a network with 79,089 edges. 

The extra edges do not have anatomical equivalents in the adult connectome, but we could consider that young worms have an expanded connectome that gets reduced in size during maturation due to pruning of neuronal synapses and axons \cite{oren2016sex,wadsworth2005axon}. Future work could explore what happens if we approximate development by slowly pruning the complete graph toward the biological connectome during self-optimization.

\subsection {Model dynamics}

Following the Hopfield network formalism, neurons have binary activation states $(-1,1)$. Asynchronous state updates are based on the following equation:

\begin{equation}\label{st_upd}
s_i(t+1) = \theta \Bigg[\sum_j^N\bigg(\sum_kw_{ijk}\bigg)s_j(t)\Bigg]
\end{equation}
where $s_i$ is the state of neuron $i$,  $w_{ijk}$ is the connection weight between neuron $i$ and neuron $j$ with index $k$ (more than one connection with the same direction could arise between $i$ and $j$), and $\theta$ is a Heaviside threshold function that returns $-1$ and $1$ for negative and positive arguments, respectively. In a Hopfield network, a node $i$ satisfies a constraint in its interaction with node $j$ with connection index $k$ if $s_is_jw_{ijk} > 0$. System energy ($E$) represents the total constraint satisfaction achieved by the network:
\begin{equation}
E = -\sum_{ijk}^Nw_{ijk}^O(t)s_i(t)s_j(t)
\end{equation}
where $w^O_{ijk}$ is the original weight configuration of $w_{ijk}$ defined at the beginning of the algorithm. In other words, if the weights of the connections between two neurons are such that they result in excitation (inhibition), the contribution towards $E$ will be negative (positive) when the states of the two neurons coincide (are different). When the self-optimization process is ongoing, Hebbian learning is applied to all system connections (i.e., change in weight, $\Delta w_{ij} = \delta s_i s_j$ , $\delta > 0$), but technically these weight changes are stored separately in another weight configuration. Keeping track of the accumulating weight changes separately in this manner allows us to plug in the state configurations produced by the learning network into $w^O_{ijk}$ to test how well these states satisfy the constraints posed by the original connectome. We fixed the learning rate ($\delta$) to 0.0001 for all simulation runs.

\begin{figure}[hbt!]
\centering
\includegraphics[width=3.2in]{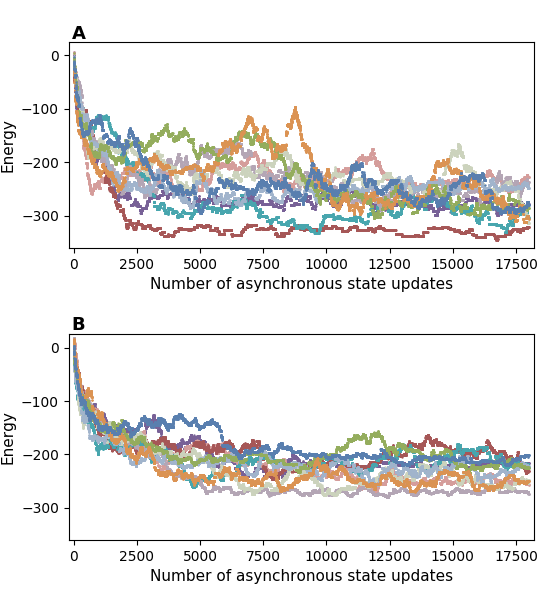}
\caption{\small{Trajectories of Hopfield network energy during neuron state updates. The plot superimposes 10 iterations of the reset-convergence cycle in the absence of Hebbian learning. Starting from arbitrary initial states the network consistently converges on lower energy, which indicates increased coordinated neural activity. In scenario \textbf{A} the neural network is initialized with inhibitory connections in accordance with the connectome's inhibitory neurons, which means that 26\% of connections are initialized with negative weights. Scenario \textbf{B} is a control condition in which the neural network is also initialized with 26\% negative connections, but these are arbitrarily distributed. Compared to this control condition, the network based on the inhibitory neurons tends to converge on lower energy values. The distribution of these final energy values is systematically compared in Fig. \ref{self_o}.}}
 \label{local}
\end{figure}

The self-optimization algorithm consists of repeating the following sequence of steps:
\begin{enumerate}
\item Arbitrary assignment of states for the neurons (reset).
\item Asynchronous updates of the neurons, typically for sufficient iterations such that activity stabilizes (convergence).
\item Application of Hebbian learning. 
\end{enumerate}

Each repetition of steps 1) and 2) is called a reset-convergence cycle. The number of neuron updates performed during step 2) is set to 18,000. We observed this to be an adequate quantity to reach stability at the end of each reset-convergence cycle. For purposes of comparison, we also measured the performance of the reset-convergence cycles before and after self-optimization (i.e., in the absence of Hebbian learning).

\begin{figure*}[hbt!]
\centering
\includegraphics[width=4in]{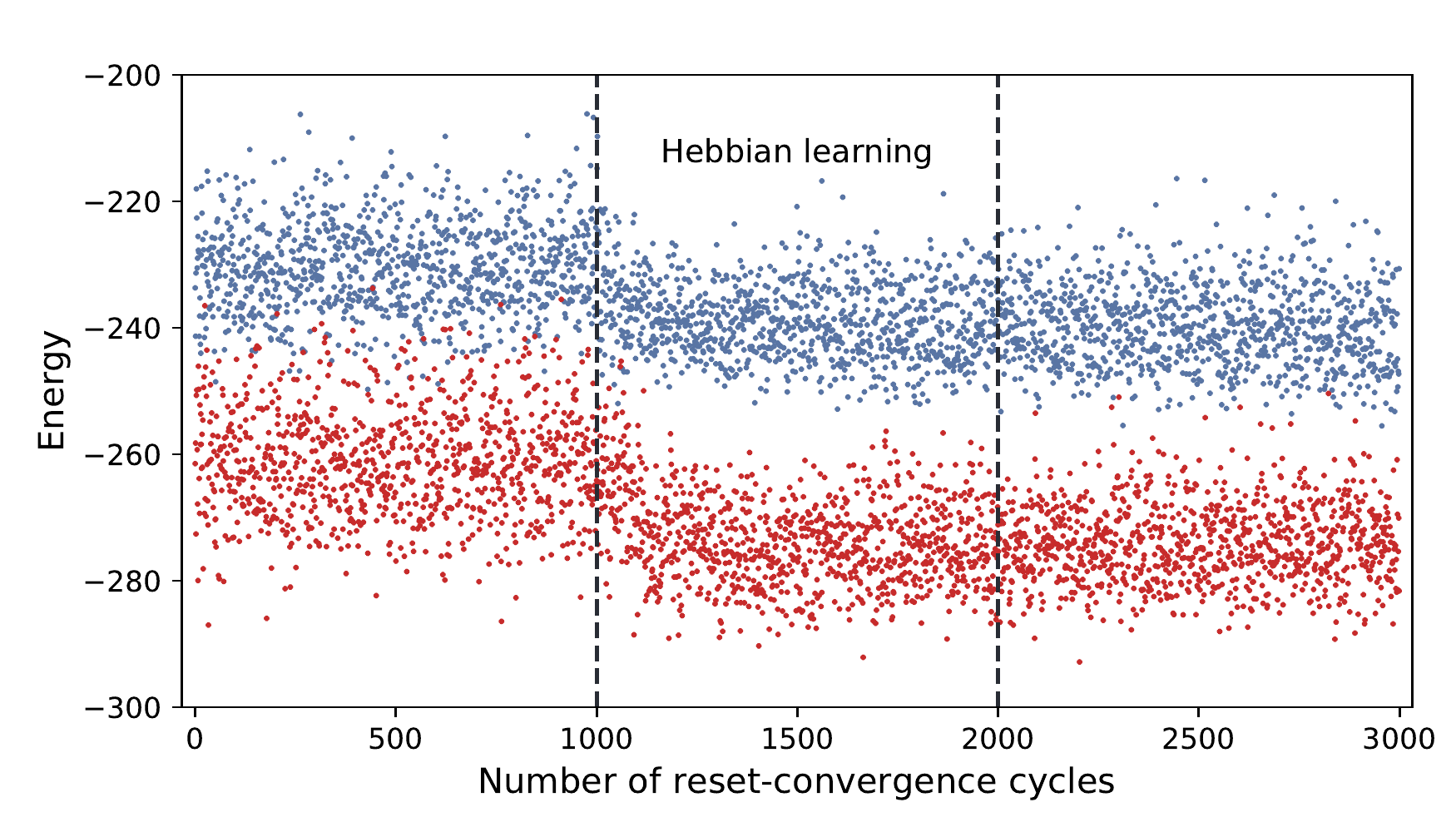}
\caption{\small{Self-optimization of neural activity in two scenarios. The plot shows a measure of neural state coordination, i.e. the total energy of a Hopfield neural network, after convergence from random initial states, across three phases for ease of comparison: 1) before learning (1–1,000), 2) during the self-optimization process based on Hebbian learning (1,001–2,000), and 3) after learning (2,001–3,000). Scenario \textbf{A (red)}: The neural network is initialized with inhibitory connections in accordance with the connectome's inhibitory neurons. This means that 26\% of connections are initialized with negative weights. Scenario \textbf{B (blue)}: Control condition in which the network is initialized with 26\% negative connections, but their location is distributed arbitrarily across the network. Each scenario was tested with 10 independent runs. Both scenarios show evidence of self-optimization in terms of spontaneous convergence on lower energy values following application of Hebbian learning. However, when inhibitory connections are distributed in accordance with the connectome's inhibitory neurons, the network tends to consistently find lower energy values, which indicates that more neuron states are coordinated.}}
\label{self_o}

\includegraphics[width=4in]{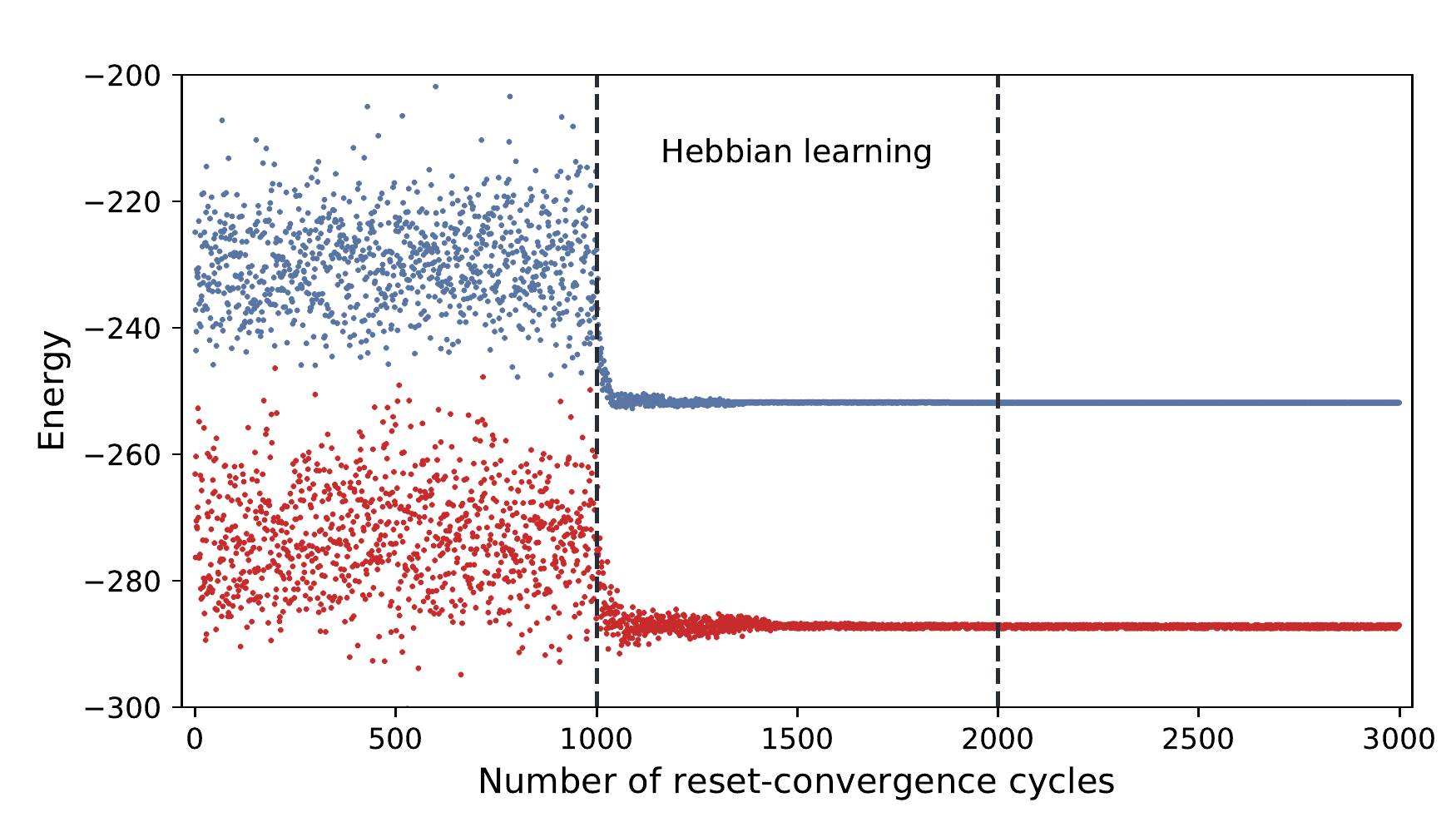}
\caption{\small{Self-optimization of coordinated neural activity in two scenarios involving excess connections.  The plot shows a measure of neural state coordination, i.e. the total energy of a Hopfield neural network, after convergence from random initial states, across three phases for ease of comparison: 1) before learning (1–1,000), 2) during the self-optimization process based on Hebbian learning (1,001–2,000), and 3) after learning (2,001–3,000). Scenario \textbf{A (red)}: The neural network is initialized with inhibitory connections in accordance with the connectome's inhibitory neurons, as in Fig. 3, but this time the network is also padded with excess connections of zero weight that turn it into a potentially fully connected network. Scenario \textbf{B (blue)}: Control condition in which the network is initialized with 26\% negative connections with an arbitrary distribution across the network, as in Fig. 3, but this time the network is also padded with excess connections of zero weight that turn it into a potentially fully connected network. Each scenario was tested with 10 independent runs. Both scenarios show evidence of self-optimization in terms of spontaneous convergence on lower energy values following application of Hebbian learning. However, when inhibitory connections are distributed in accordance with the connectome's inhibitory neurons, the network tends to consistently find lower energy values, which indicates that more neuron states are coordinated. The inclusion of excess connections made self-optimization significantly more effective.}}
\label{self_o_extra_z}
\end{figure*}

\section{Results}

We first explored the reset-convergence dynamics of the Hopfield network of the connectome without Hebbian learning. Fig. \ref{local} shows network energy after successive neuron state updates starting from arbitrary initial states. Energy tends to decrease, but the network does not necessarily reach a fixed-point attractor because the presence of inhibitory synapses makes neural coordination more difficult, as we had already previously observed \cite{morales2019alife,morales2020unsupervised}. We noted a tendency for the biological distribution of inhibitory connections to facilitate convergence on more coordinated states.

We then investigated this network's capacity for self-optimization more systematically. First, we performed 1000 reset-convergence cycles without Hebbian learning. Then, we applied self-optimization using 1000 reset-convergence cycles. Finally, we applied to the network 1000 reset-convergence cycles without Hebbian learning using configuration obtained by self-optimization to show its stability. Cycles 1001-2000 follow steps 1, 2, and 3 of the self-optimization algorithm. Cycles 1-1000 and 2001-3000 only follow steps 1 and 2 because they are not learning phases. Results are averaged over 10 different experiments with a different initial random number seed. We performed this set of simulation runs using the network with the biological distribution of inhibitory connections (scenario A), and repeated the simulation runs using the network with an arbitrary distribution of inhibitory connections (scenario B). 

The results are shown in Fig. \ref{self_o}, which reveals consistently improved neural coordination in scenario A compared to scenario B. In addition, we find that Hebbian learning manages to improve the performance of both networks, thus demonstrating that self-optimization can be applied to them. However, the resulting improvement in neural coordination following self-optimization is not as pronounced as in our previous work. Therefore, we also tested these networks as complete graphs, which indeed significantly improves self-optimization, as shown in Fig. \ref{self_o_extra_z}.

\section{Discussion}

We have tested self-optimization in a Hopfield network based on the \textit{C. elegans} connectome. Here we suggest some possible implications of our work for theoretical neuroscience. Of course, we acknowledge that our model is highly abstract, and that the link with actual brain function is not direct; so we must be cautious. Nevertheless, there are some interesting considerations that can be made. That the network starts with a large amount of different initial attractors can be interpreted in terms of the nervous system being initially pre-configured with a large number of different cell assembly configurations, which then get optimized according to lifetime demands placed on them \cite{gyorgy2019brain}. The reset-convergence cycle could be interpreted in terms of biological rhythms that modulate neural activity, similar to sleep-wake cycles \cite{woodward2015neural}. 

We showed that by assigning inhibitory connections according to neuron identity, the network tends to converge on lower energy values from arbitrary initial states. This implies that the distribution of inhibitory neurons in the connectome plays a role in facilitating neural coordination. We also demonstrated that when we repeat such reset-convergence cycles in the presence of accumulating weight changes due to Hebbian learning there is a consistent further decrease in energy values. In other words, self-optimization additionally facilitates coordination among neurons. However, we do not find that the improvement generated by self-optimization is significantly facilitate by a realistic distribution of inhibitory connections; the relative decrease in energy values is comparable across the scenarios based on biological and arbitrary distributions.

Adding extra zero-weighted connections to the initial networks notably improves the performance of self-optimization, which then tends to focus the state space of the networks to the most coordinated state configurations. This difference in performance deserves more attention in future work, and the biological equivalent of the excess connections remains to be determined. We mentioned the possibility that an expanded network could correspond to the connectome of an immature worm before the pruning of synapses and axons. Another possibility is that these connections could model potential influences between neurons that go beyond synaptic connections in the adult worm, for example direct neurotransmitter diffusion between nearby neurons. For this we could draw on the literature on chemical signaling in the worm's nervous system \cite{li99neuropeptide}, and combine it with relevant artificial neural network architectures \cite{husbands2010spatial}. 

This improvement of biological realism of our simulation model could go hand in hand with the implementation of more realistic neural dynamics, for instance by using continuous-time, continuous-state neurons like in \cite{zarco2018continous}. Finally, we would ultimately need to embody this network into a simulated worm body in order to test it as a coupled agent-environment system \cite{izquierdo2008analysis}, which would allow us to test if neural self-optimization improves behavioral coordination.

\bibliographystyle{abbrv}
\bibliography{inhib}

\begin{thebibliography}{10}

\bibitem{aguilera2017}
M.~Aguilera, C.~Alqu{\'e}zar, and E.~J. Izquierdo.
\newblock Signatures of criticality in a maximum entropy model of the
  \textit{C. elegans} brain during free behaviour.
\newblock In C.~Knibbe, G.~Beslon, D.~Parsons, D.~Misevic, J.~Rouzaud-Cornabas,
  N.~Bred\`{e}che, S.~Hassas, O.~Simonin, and H.~Soula, editors, {\em The
  Fourteenth European Conference on Artificial Life}, pages 29--35. MIT Press,
  2017.

\bibitem{albertson1976pharynx}
D.~G. Albertson and J.~Thompson.
\newblock The pharynx of \textit{Caenorhabditis elegans}.
\newblock {\em Philosophical Transactions of the Royal Society of London.
  Biological Sciences}, 275(938):299--325, 1976.

\bibitem{anderson2010celegans}
J.~L. Anderson, L.~T. Morran, and P.~C. Phillips.
\newblock Outcrossing and the maintenance of males within \textit{C. elegans}
  populations.
\newblock {\em Journal of Heredity}, 101(suppl\_1):S62--S74, 2010.

\bibitem{beer2016propagation}
R.~D. Beer and E.~J. Izquierdo.
\newblock Propagation of rhythmic dorsoventral wave in a neuromechanical model
  of locomotion in \textit{Caernohabditis elegans}.
\newblock In C.~Gershenson, T.~Froese, J.~Siqueiros-Garc\'{i}a, W.~Aguilar, and
  H.~Sayama, editors, {\em Proceedings of the Artificial Life Conference 2016},
  pages 544--545. MIT Press, 2016.

\bibitem{brunel2000}
N.~Brunel.
\newblock Dynamics of sparsely connected networks of excitatory and inhibitory
  spiking neurons.
\newblock {\em Journal of Computational Neuroscience}, 8(3):183--208, 2000.

\bibitem{buxhoeveden2002}
D.~P. Buxhoeveden and M.~F. Casanova.
\newblock The minicolumn hypothesis in neuroscience.
\newblock {\em Brain}, 125(5):935--951, 2002.

\bibitem{cook2019whole}
S.~J. Cook et~al.
\newblock Whole-animal connectomes of both \textit{Caenorhabditis elegans}
  sexes.
\newblock {\em Nature}, 571(7763):63--71, 2019.

\bibitem{dahlberg2020}
B.~A. Dahlberg and E.~J. Izquierdo.
\newblock Contributions from parallel strategies for spatial orientation in
  \textit{C. elegans}.
\newblock In J.~Bongard, J.~Lovato, L.~Hebert-Dufr\'{e}sne, R.~Dasari, and
  L.~Soros, editors, {\em The 2020 Conference on Artificial Life}, pages
  16--24. MIT Press, 2020.

\bibitem{worm2006review}
L.~R. Girard et~al.
\newblock Wormbook: the online review of \textit{Caenorhabditis elegans}
  biology.
\newblock {\em Nucleic Acids Research}, 35:D472--D475, 2006.
\newblock doi:10.1093/nar/gkl894.

\bibitem{gyorgy2019brain}
M.~Gy{\"o}rgy~Buzs{\'a}ki.
\newblock {\em The brain from inside out}.
\newblock Oxford University Press, 2019.

\bibitem{hattori2012modeling}
Y.~Hattori, M.~Suzuki, Z.~Soh, Y.~Kobayashi, and T.~Tsuji.
\newblock Modeling of the pharyngeal muscle in \textit{Caenorhabditis elegans}
  based on {FitzHugh-Nagumo} equations.
\newblock {\em Artificial Life and Robotics}, 17(2):173--179, 2012.

\bibitem{hendry1987}
S.~H. Hendry, H.~Schwark, E.~Jones, and J.~Yan.
\newblock Numbers and proportions of {GABA-immunoreactive} neurons in different
  areas of monkey cerebral cortex.
\newblock {\em Journal of Neuroscience}, 7(5):1503--1519, 1987.

\bibitem{husbands2010spatial}
P.~Husbands, A.~Philippides, P.~Vargas, C.~L. Buckley, P.~Fine, E.~Di~Paolo,
  and M.~O'Shea.
\newblock Spatial, temporal, and modulatory factors affecting {GasNet}
  evolvability in a visually guided robotics task.
\newblock {\em Complexity}, 16(2):35--44, 2010.

\bibitem{izquierdo2008analysis}
E.~Izquierdo and T.~B{\"u}hrmann.
\newblock Analysis of a dynamical recurrent neural network evolved for two
  qualitatively different tasks: Walking and chemotaxis.
\newblock In {\em ALIFE}, pages 257--264, 2008.

\bibitem{izquierdo2018role}
E.~J. Izquierdo.
\newblock Role of simulation models in understanding the generation of behavior
  in \textit{C. elegans}.
\newblock {\em Current Opinion in Systems Biology}, 13:93--101, 2018.

\bibitem{jarrell2012connectome}
T.~Jarrell et~al.
\newblock The connectome of a decision-making neural network.
\newblock {\em Science}, 337(6093):437--444, 2012.

\bibitem{li99neuropeptide}
C.~Li, L.~S. Nelson, K.~Kim, A.~Nathoo, and A.~C. Hart.
\newblock Neuropeptide gene families in the nematode \textit{Caenorhabditis
  elegans}.
\newblock {\em Annals of the New York Academy of Sciences}, 897(1):239--252,
  1999.

\bibitem{morales2019alife}
A.~Morales and T.~Froese.
\newblock Self-optimization in a hopfield neural network based on the
  \textit{C. elegans} connectome.
\newblock In H.~Fellerman, J.~Bacardit, A.~Goñi-Moreno, and R.~Füchslin,
  editors, {\em The 2019 Conference on Artificial Life}, pages 448--453. MIT
  Press, 2019.

\bibitem{morales2020unsupervised}
A.~Morales and T.~Froese.
\newblock Unsupervised learning facilitates neural coordination across the
  functional clusters of the \textit{C. elegans} connectome.
\newblock {\em Frontiers in Robotics and AI}, 7:40, 2020.

\bibitem{oren2016sex}
M.~Oren-Suissa, E.~A. Bayer, and O.~Hobert.
\newblock Sex-specific pruning of neuronal synapses in \textit{Caenorhabditis
  elegans}.
\newblock {\em Nature}, 533(7602):206--211, 2016.

\bibitem{pereira2015}
L.~Pereira et~al.
\newblock A cellular and regulatory map of the cholinergic nervous system of
  \textit{C. elegans}.
\newblock {\em Elife}, 4:e12432, 2015.

\bibitem{riddle97neurotr}
D.~Riddle.
\newblock Neurotransmitter metabolism and function.
\newblock In D.~Riddle, T.~Blumenthal, and B.~Meyer, editors, {\em C. elegans
  II}. Cold Spring Harbor Laboratory Press, New York, 1997.

\bibitem{sanyal2004}
S.~Sanyal et~al.
\newblock Dopamine modulates the plasticity of mechanosensory responses in
  \textit{Caenorhabditis elegans}.
\newblock {\em The EMBO journal}, 23(2):473--482, 2004.

\bibitem{sawin2000}
E.~R. Sawin, R.~Ranganathan, and H.~R. Horvitz.
\newblock \textit{C. elegans} locomotory rate is modulated by the environment
  through a dopaminergic pathway and by experience through a serotonergic
  pathway.
\newblock {\em Neuron}, 26(3):619--631, 2000.

\bibitem{sohn2011}
Y.~Sohn, M.-K. Choi, , Y.-Y. Ahn, J.~Lee, and J.~Jeong.
\newblock Topological cluster analysis reveals the systemic organization of the
  \textit{Caenorhabditis elegans} connectome.
\newblock {\em PLOS Computational Biology}, 7(5), 2011.

\bibitem{triplett2018}
M.~A. Triplett, L.~Avitan, and G.~J. Goodhill.
\newblock Emergence of spontaneous assembly activity in developing neural
  networks without afferent input.
\newblock {\em PLoS computational biology}, 14(9):1--22, 2018.

\bibitem{wadsworth2005axon}
W.~G. Wadsworth.
\newblock Axon pruning: \textit{C. elegans} makes the cut.
\newblock {\em Current biology}, 15(19):R796--R798, 2005.

\bibitem{watson2011optimization}
R.~A. Watson, C.~L. Buckley, and R.~Mills.
\newblock Optimization in “self-modeling” complex adaptive systems.
\newblock {\em Complexity}, 16(5):17--26, 2011.

\bibitem{watson2011global}
R.~A. Watson, R.~Mills, and C.~L. Buckley.
\newblock Global adaptation in networks of selfish components: Emergent
  associative memory at the system scale.
\newblock {\em Artificial Life}, 17(3):147--166, 2011.

\bibitem{watson2011transformations}
R.~A. Watson, R.~Mills, and C.~L. Buckley.
\newblock Transformations in the scale of behavior and the global optimization
  of constraints in adaptive networks.
\newblock {\em Adaptive Behavior}, 19(4):227--249, 2011.

\bibitem{white1986structure}
J.~G. White, E.~Southgate, J.~N. Thomson, and S.~Brenner.
\newblock The structure of the nervous system of the nematode
  \textit{Caenorhabditis elegans}.
\newblock {\em Philosophical Transactions of the Royal Society, Biological
  Sciences}, 314(1165):1--340, 1986.
\newblock doi:10.1098/rstb.1986.0056.

\bibitem{winkler2009}
D.~A. Winkler, F.~R. Burden, and J.~D. Halley.
\newblock Predictive mesoscale network model of cell fate decisions during
  \textit{C. elegans} embryogenesis.
\newblock {\em Artificial Life}, 15(4):411--421, 2009.

\bibitem{woodward2015neural}
A.~Woodward, T.~Froese, and T.~Ikegami.
\newblock Neural coordination can be enhanced by occasional interruption of
  normal firing patterns: A self-optimizing spiking neural network model.
\newblock {\em Neural Networks}, 62:39 -- 46, 2015.

\bibitem{zarco2018self}
M.~Zarco and T.~Froese.
\newblock Self-modeling in {Hopfield} neural networks with continuous
  activation function.
\newblock {\em Procedia {Computer Science}}, 123:573--578, 2018.

\bibitem{zarco2018continous}
M.~Zarco and T.~Froese.
\newblock Self-optimization in continuous-time recurrent neural networks.
\newblock {\em Frontiers in Robotics and AI}, 5:96, 2018.

\end{thebibliography}

\end{document}